\documentclass[MSNbibl,nameyear,dvips]{arxstspdf}
\usepackage{flushend}
\usepackage{stfloats}
\usepackage{graphicx}

\volume{29}
\issue{3}
\pubyear{2014}
\firstpage{420}
\lastpage{424}
\doi{10.1214/14-STS468} 


\begin{document}
\begin{frontmatter}

\title{Revisiting Francis Galton's Forecasting Competition}
\runtitle{Revisiting Francis Galton's Forecasting Competition}

\begin{aug}
\author[a]{\fnms{Kenneth F.}~\snm{Wallis}\corref{}\ead[label=e1]{K.F.Wallis@warwick.ac.uk}}
\runauthor{K. F. Wallis}

\affiliation{University of Warwick}

\address[a]{Kenneth F. Wallis is Emeritus Professor of Econometrics,
Department of Economics, University of Warwick,
Coventry, CV4 7AL, United Kingdom \printead{e1}.}

\end{aug}

\begin{abstract}
This note reexamines the data from a weight-judging
competition described in an article by Francis Galton published in 1907.
Following the correction of some errors, it is shown that this forecasting
competition is an interesting precursor of two more recent developments in
the statistical forecasting literature. One is forecast combination, with
the mean forecast here exactly coinciding with the outcome, and the second
is the use of two-piece frequency and probability distributions to describe
asymmetry.
\end{abstract}


\end{frontmatter}

\section{Introduction: The Weight-Judging Competition}\label{sec1}

In an article titled \textit{Vox Populi} in the weekly journal of science,
\textit{Nature}, of March 7, 1907, Francis Galton, renowned anthropologist,
biometrician and statistician, described a weight-judging competition
conducted at the recent West of England Fat Stock and Poultry Exhibition.
``A fat ox having been selected, competitors bought stamped and numbered
cards, for 6\textit{d.} each, on which to inscribe their respective names,
addresses, and estimates of what the ox would weigh after it had been
slaughtered and `dressed.' Those who guessed most successfully received
prizes.'' In accordance with his lifelong motto, reported by his biographer
Karl Pearson as ``Whenever you can, count'' (Pearson, \citeyear{11}, page~340), Galton was able to
obtain the loan of all the entry cards for a short period. On studying the
entries, he found that ``these afforded excellent material. The judgments
were unbiassed by passion \dots. The sixpenny fee deterred practical
joking, and the hope of a prize and the joy of competition prompted each
competitor to do his best. The competitors included butchers and farmers,
some of whom were highly expert in judging the weight of cattle.''

After weeding out 13 defective or illegible cards, there were 787 entries
for analysis, and Galton began by constructing a ranked list of the
estimates, converting from hundredweights, quarters and pounds to pounds.
He then picked out the 5, 10, 15, \dots, 95 percentiles, which are
tabulated as the ``distribution of the estimates'' in his article, following
his ``method of percentiles,'' developed over many years. His preferred
measure of central tendency is the median, although, having earlier
introduced this term, in this article he retains the previous terminology:
``According to the democratic principle of `one vote one value,' the
middlemost estimate expresses the \textit{vox populi}, every other estimate
being condemned as too low or too high by a majority of the voters.'' He
reports that ``the middlemost estimate is 1207 lb., and the weight of the
dressed ox proved to be 1198 lb.; so the \textit{vox populi} was in this
case 9 lb., or 0.8 per cent. of the whole weight too high.'' Galton
concludes that ``This result is, I think, more creditable to the
trustworthiness of a democratic judgment than might have been expected.''

Such a competition provides an early example of the forecast competitions
that have become familiar in the forecasting literature. In this note we
take another look at Galton's data, and show that his article, despite some
inaccuracy, is an interesting precursor of two more recent developments in
the statistical forecasting literature. It has also attracted attention in
the public choice literature; see, for example, \citet{10}, who
reproduce the original article (\cite{6}), together with a Letter to
the Editor of \textit{Nature} published a week earlier, arguing for the
median on democratic principles (\cite{5}). Also, more popularly, see
Surowiecki's \textit{The Wisdom of Crowds} (\citeyear{15}), which begins with the
weight-judging competition and so brings it to the attention of a wider
audience.

\section{A Visit to the Archives: Some~Discrepancies}\label{sec2}

Galton's working papers, notes, some correspondence and a handwritten draft
of his article are stored in the Galton Archive at University College,
London. Study of this material reveals some slips that have a bearing on
subsequent analysis and interpretation. Galton reached the age of 85 on
February 16, 1907, and may have been in a hurry to attract ``immediate
attention,'' for which \textit{Nature} provided ``a ready means'' of
communication, as Pearson observed (Pearson, \citeyear{11}, page 400).

\begin{figure}

\includegraphics{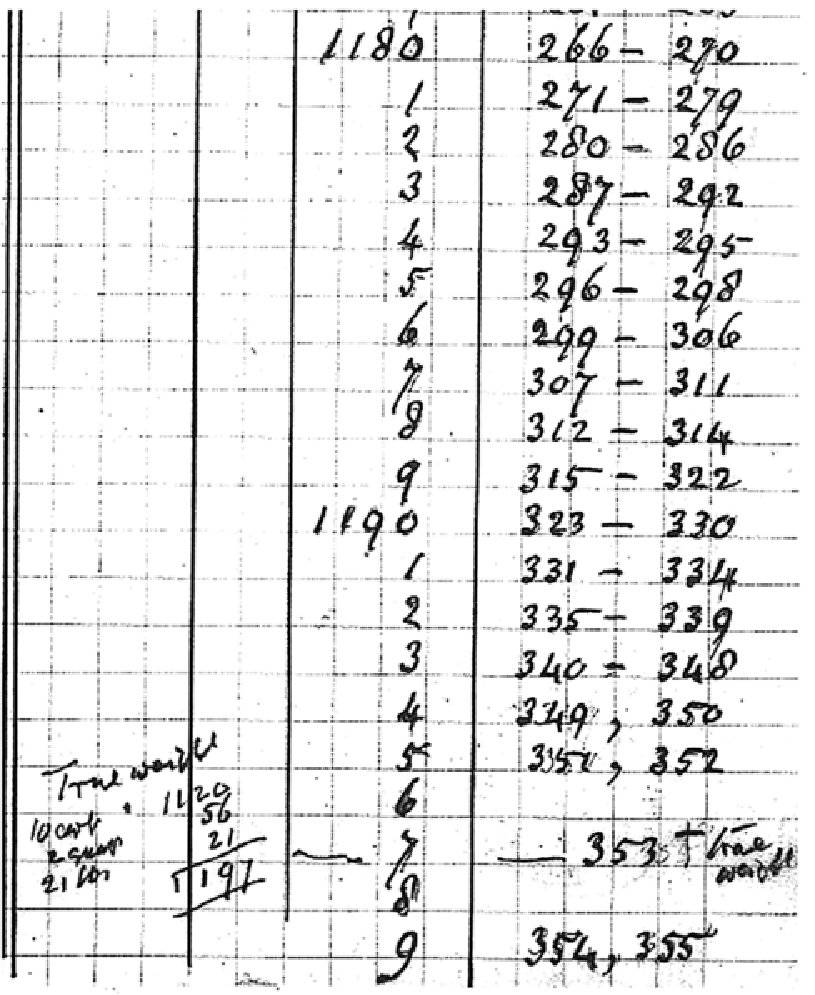}

\caption{Extract from Galton's worksheets.
Galton recorded the rank order(s) of
the competitor(s) who chose each possible weight, recorded in pounds. In
this relatively dense part of the distribution of estimated weights, it is
remarkable that there was only a single winner, and no immediately adjacent
runner-up.}\label{fig1}
\end{figure}

There are small errors in all three figures appearing in the summary
statement of the results quoted above, although they are arithmetically
consistent. First, with respect to the median entry, among 787 observations
this is the 394{th} in the ranked list, which is 1208 pounds. Second,
the outcome, that is, the dressed weight of the ox, was reported in a
letter from the organiser of the competition as 10 cwt, 2 qt and 21~lbs,
that is, 1197 pounds, and this figure appears in Galton's worksheets, being
equal to the 353{rd} entry in the ranked list, as shown in the extract
reproduced in Figure~\ref{fig1}. So the true error in the middlemost estimate is 11~lb.
Galton devoted a paragraph of a letter to his nephew, Edward Wheler
Galton, dated February 4, 1907, to this subject. He says that he is ``just
now at some statistics that might interest you,'' and concludes a brief
account of the weight-judging data with the statement ``The average was 11
lbs. wrong'' (\cite*{12}, page 581).

Galton's handwritten draft of the article, however, presents the results as
follows. ``The weight of the dressed ox proved to be 1198 lbs. The estimates
were scattered about their own middlemost value of 1208 lbs so the vox
populi was in this case 11 lbs too high, or closely 1\% of the weight.''
Here the arithmetic is inconsistent, because the outcome has been
incorrectly transcribed. But instead of correcting 1198 to 1197, so that
the difference is 11, as correctly stated in the draft, in the published
version 1208 has been changed to 1207, and the error is correspondingly
reported as 9 lb. In attempting the correction, the wrong four-digit number
has had its final 8 altered to 7. It is not clear at what point in the work
this happened, since the wrong median also appears in the published table
of the distribution of the estimates, although the first and third
quartiles are correctly given. With 787 observations these require no
interpolation, being the 197{th} and 591{st} observations in the
ranked list, equal to 1162 and 1236 respectively.

Galton appears to have remained unaware that the results as presented in
his article were inaccurate, since they reappear in his \textit{Memories of
My Life} (\citeyear{8}). A~short account of his visit to the cattle exhibition ``a
little more than a year ago'' and his subsequent research is given (pages
280--281); this refers to his ``memoir published in \textit{Nature}'' and
repeats the incorrect numbers to be found there, respectively 1207 pounds
and 1198 pounds.

\section{Forecast Combination}\label{sec3}

The idea that combining different forecasts of the same event might be
worthwhile has gained wide acceptance since the seminal article of
\citet{2}, some sixty years after Galton's \textit{Vox Populi}. A
substantial literature has subsequently appeared, mostly concerning point
forecasts of the future realisation of a random variable and, although the
median of a set of competing forecasts is sometimes a combined forecast of
interest, simple averages and various weighted averages are more common,
given that the statistical forecasting literature is largely founded on
least squares principles.

The choice between the median and the mean was discussed in a letter to the
editor of \textit{Nature} published two weeks after Galton's article had
appeared. The correspondent, \citet{9}, wished that Galton had also
reported the arithmetic mean of the 787 observations. He says that ``I have
not the actual figures, but judging from the data in Mr. Galton's article,
the mean would seem to be approximately 1196 lb., which is much closer to
the ascertained weight (1198 lb.) than the median (1207 lb.)'': he had
calculated the mean of the percentiles in Galton's table. In his reply, one
week later, \citet{7} reports the correct mean of all the figures as
1197 pounds, which for him shows that ``the compactness of a table of
centiles is no hindrance to their wider use.'' He does not remark on the
fact that this is closer to the true value than the median, as Hooker had
observed. Indeed, using the correct value of the outcome, the mean estimate
has zero error. This early example of the gains from forecast combination
would very likely have been cited by \citet{2}, had they
been aware of it.

Surowiecki focusses on this mean estimate, reported in Galton's reply, as
representative of ``the collective wisdom of the Plymouth crowd''
(Suro\-wiecki, \citeyear{15}, page xiii). He compares it to the published outcome of 1198 pounds and concludes
that ``the crowd's judgment was essentially perfect.'' Had he known the true
outcome, his conclusion on the wisdom of crowds would no doubt have been
more strongly expressed.

In this one-off exercise there is no possibility of using calculated
weights to construct an alternative combined forecast. Other possibilities
considered in the forecast combination literature include trimmed means: in
this case symmetric trimming of the sample by increasing amounts simply
moves the estimate from the mean of 1197 towards the median, which is the
extreme example of a trimmed mean, and increases the error.

\section{Two-Piece Distributions}\label{sec4}

\begin{figure*}

\includegraphics{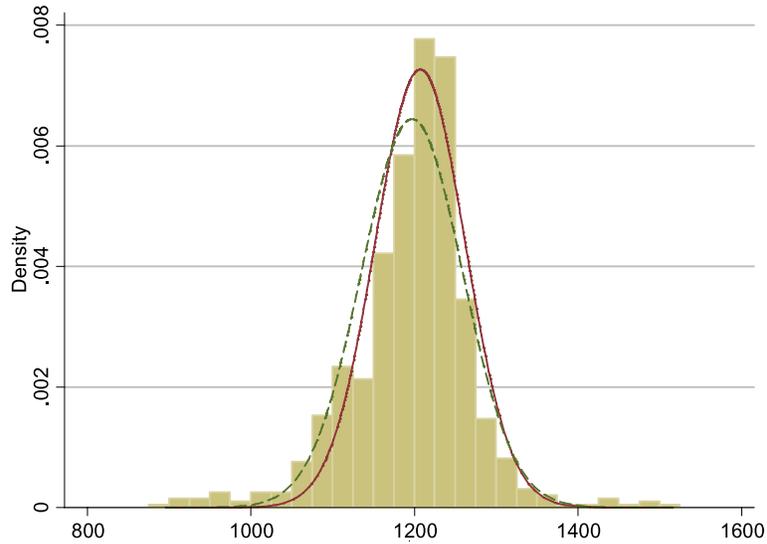}

\caption{Fitted normal curves.
Histogram: original data.
Solid line: PDF of the $N(1207, 54.9)$ distribution (Galton).
Dashed line: PDF of the $N(1197, 61.9)$ distribution (Pearson).}\label{fig2}
\end{figure*}

Galton discusses the dispersion and shape of the distribution of the
individual forecasts of the dressed weight with reference to the normal
distribution. As a measure of spread he uses the ``probable error,'' an
archaic term for the semi-interquartile range: the probability of obtaining
a value within one probable error of the centre of a symmetric distribution
is 0.5. For normal variables the probable error is equal to 0.6745 $\times$
(standard deviation), hence, Galton's probable error of $(1236 -
1162) /2 = 37$ is equivalent to a standard deviation of 54.9. To
compare the empirical distribution with a normal distribution with mean
($=$ median) 1207 and probable error 37, he plots each distribution's
percentiles against the corresponding percentages (5, 10, \dots, 95); his
``ogive'' curves are now called inverse cumulative distribution functions.
His diagram shows that the normal distribution does not extend sufficiently
far into the tails of the empirical distribution. To modern eyes, this is
more readily apparent in the comparison of the probability density function
(PDF) of Galton's normal distribution and the sample histogram shown in
Figure~\ref{fig2}.

In the course of an extended review of Galton's statistical investigations,
Pearson (\citeyear{11}), Chapter~13, revisits this data set. In respect of Galton's
methods, he observes (page 400) ``That Galton used median and quartiles so
frequently \dots must, I think, be attributed to his great love of brief
analysis. He found arithmetic in itself irksome;'' For the present exercise,
he notes (page 404) that ``the percentile method of tabulation does not
permit of very ready determination of the mean and standard deviation and
so of getting the best normal distribution,'' in other words, the one fitted
by his method of moments! He finds, ``after some labour: mean 1197, standard
deviation 61.895, \dots. These give a far better fit than Galton's median
and quartile values.'' Pearson seems to have overlooked the mean value given
by \citet{7}. Figure~\ref{fig2} shows that his claim of ``a far better fit'' is
true away from the centre of the distribution, but less true around the
peak.

A major problem, as Galton observes, is that the distribution of forecasts
is asymmetric, the extent of underestimation of the outcome being the
greater. His final suggestion is that the lower half of the distribution
might agree well with a normal distribution with probable error 45, and the
upper half with one having probable error 29. He had made a similar
suggestion on a previous occasion (\cite{4}), only to have it
immediately pointed out that placing two half-normal distributions together
in this way would result in a discontinuity at the join point (\cite{17}).
The two-piece normal distribution, introduced by \citet{3} and
subsequently rediscovered several times (Wallis, \citeyear{16}), appropriately
scales each half-normal distribution and is continuous at the mode. This
distribution has found widespread use as a representation of asymmetric
risks in density forecasting.

A second major problem, however, is
that the sample distribution is leptokurtic, that is, more peaked and with
fatter tails than the normal distribution: the conventional $\beta_{2}$  measure of
kurtosis is 6.01. The corresponding lack of ``shoulders'' in the distribution
explains why Galton's estimate of the standard deviation based on the
interquartile range is smaller than Pearson's second moment estimate. In
the two-piece normal distribution we have $3 \leq \beta_2 \leq 3.87$, but
this apparent kurtosis is simply a consequence of its asymmetry, with each
half-normal distribution being rescaled, not reshaped. Attention then turns
to two-piece versions of more kurtotic distributions, such as the
Student-$t$ distribution with a relatively small degrees of freedom.
However, incorporating skewness in this way mainly accommodates skewness in
the central part of the distribution, and does not allow the two tails of
the distribution to have different rates of decay. Accordingly, \citet{18} develop a class of generalized asymmetric Student-$t$
(AST) distributions which has one skewness parameter and two tail
parameters, which offer the possibility of improved fit in the tail
regions.

\begin{figure*}

\includegraphics{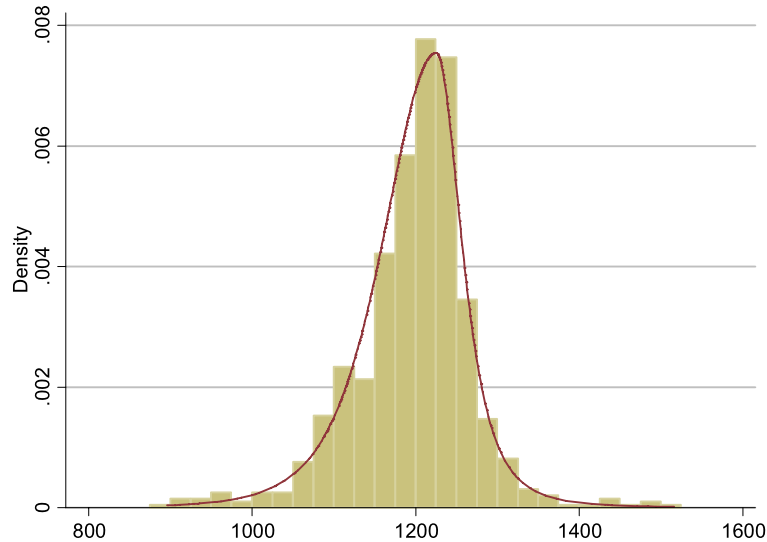}

\caption{The asymmetric Student-$t$ distribution.
Histogram: original data.
Solid line: PDF of the fitted AST distribution (Zhu and Galbraith, \citeyear{18}).}\label{fig3}
\end{figure*}

Fitting the (CDF of the) AST distribution to Galton's data by
maximum likelihood gives the result shown (as the PDF) in Figure~\ref{fig3}. It is
seen that the main features of the data are well represented by this
five-parameter form, compared to the two-parameter forms shown in Figure~\ref{fig2}.
Basing model selection on the Akaike information criterion leads to a clear
preference for the five-parameter form over any restricted version of the
distribution. The two ``degrees of freedom'' parameter estimates are 4.97
and 2.73 respectively in the lower and upper pieces of the distribution.
The skewness measure of \citet{1}, which is equal to the
difference between the areas under the PDF above and below the mode, is
$-0.32$. It is hoped that this successful example will encourage further
applications of the AST distribution, not only in financial econometrics,
for which it was designed, but also in other fields.

\section{Epilogue}\label{sec5}

Nothing in this account of Galton's inaccuracies should be taken as
detrimental to his standing as a major figure in the development of modern
statistics. Stigler (\citeyear{13}), Chapter~8, assesses his contributions,
describing him (page 266) as ``a romantic figure in the history of
statistics, perhaps the last of the gentleman scientists.'' More recently
(\cite{14}), he dates the beginning of the half century that he calls
the Statistical Enlightenment from Galton's address to the 1885 meeting of
the British Association for the Advancement of Science. By 1907, however,
Galton's major accomplishments were behind him, and the analysis of the
weight-judging data was the last piece of statistical work that he
published. He was aged 85, and in poor health, which was the reason for his
presence in the West of England. He explained in a letter to his nephew in
October 1906 that ``London in November would help to, or quite, kill me,''
but whereas it had been his custom to spend winters in Southern Europe, ``I
funk now foreign travel'' (\cite*{12}, page 579). He had not lost his
mental powers, however, and, once in Plymouth, could not resist the
opportunity that access to these data offered him. Galton died on January
17, 1911, and is buried at Claverdon, a few miles from Warwick.

\section*{Acknowledgments}
My thanks to Gianna Boero, John Galbraith,
Federico Lampis, Jeremy Smith, two referees and the Special Collections
Librarians at University College, London.
\textit{Note}: facsimiles of the Galton and Pearson items are available at
\url{http://www.galton.org/}.



\end{document}